\documentclass[aps,prd,amssymb,superscriptaddress,onecolumn,nofootinbib,preprintnumbers]{revtex4-2}
\usepackage[utf8]{inputenc}
\usepackage[english]{babel}
\usepackage{amsmath}
\usepackage{amsfonts}
\usepackage{amssymb}
\usepackage{amsthm}
\usepackage{xcolor}
\usepackage{scalerel}
\usepackage{graphicx}
\usepackage[force]{feynmp-auto}
\usepackage{feynmp}
\usepackage[hidelinks]{hyperref}
\usepackage{mathrsfs}
\newcommand\bw{\begin{widetext}}
\newcommand\ew{\end{widetext}}

 \def\be{\begin{equation}}
\def\ee{\end{equation}}
 \def\ba{\begin{align}}
\def\ea{\end{align}}
\def\bea{\begin{eqnarray}}
\def\eea{\end{eqnarray}}

\def\a{\alpha}
\def\b{\beta}

\def\m{\mu}
\def\n{\nu}

\begin{document}
\title{{\bf Gravitational Tunneling in Lorentz Violating Gravity}}

\author{F. Del Porro}
\email[]{fdelporr@sissa.it}
\address{SISSA, Via Bonomea 265, 34136 Trieste, Italy and INFN Sezione di Trieste}
\address{IFPU - Institute for Fundamental Physics of the Universe, Via Beirut 2, 34014 Trieste, Italy}

\author{M. Herrero-Valea}
\email[]{mherrero@ifae.es}
\address{Institut de Fisica d?Altes Energies (IFAE), The Barcelona Institute of Science and Technology, Campus UAB, 08193 Bellaterra (Barcelona) Spain}

\author{S. Liberati}
\email[]{liberati@sissa.it}

\author{M. Schneider}
\email[]{mschneid@sissa.it}

\address{SISSA, Via Bonomea 265, 34136 Trieste, Italy and INFN Sezione di Trieste}
\address{IFPU - Institute for Fundamental Physics of the Universe, Via Beirut 2, 34014 Trieste, Italy}
\date{\today}
\begin{abstract}
Black holes in Lorentz violating gravity, such as Einstein--Aether or Ho\v rava--Lifshitz Gravity, are drastically different from their general relativistic siblings. Although they allow for superluminal motion in their vicinity, they still exhibit an absolute causal boundary in the form of a \emph{universal horizon}. By working in the tunneling picture for a gravitating scalar field, we show that universal horizons emit Hawking radiation in a manner akin to standard results in General Relativity, with a temperature controlled by the high-energy behavior of the dispersion relation of the gravitating field, and in agreement with alternative derivations in the literature. Our results substantiate the link between the universal horizon and thermodynamics in Lorentz violating theories.
\end{abstract}

\maketitle
\section{Introduction}
The thermodynamical features of horizons have been extensively studied ever since Hawking predicted that black holes create particles in their vicinity \cite{haw75}. His seminal work inspired a wealth of research aiming at understanding the occurrence of quantum processes at causal boundaries. Although horizons act classically as semipermeable membranes, allowing only a one-way crossing, quantum effects might follow classically forbidden processes and exit the region trapped by the horizon. Hawking tentatively described particle production as a consequence of this through quantum tunneling, but it took twenty years until such description was made manifest \cite{par00,pad02}. In this picture, particles escape the causal enclosure of the horizon through gravitational tunneling on complex paths. The formalism proposed by \cite{par00,pad02} was substantiated by \cite{mas00}, and finally connected to the Hamilton--Jacobi formalism by \cite{di07,van11}, where thermodynamical properties were shown to be in agreement with previous results that use Bogolubov coefficients. In contrast to the latter, the Hamilton--Jacobi method links the pole structure of the semi-classical amplitude at the horizon to the imaginary part of the classical action, such that the tunneling rate can be compared with the Boltzmann distribution, from which thermal properties follow immediately.

The prevalence of horizon thermodynamics underlines how deeply this effect is enmeshed in the foundation of quantum field theory. Although seemingly simple, the tunneling picture offers various insights into the local quantum processes that govern the vicinity of the horizon. Especially after its application to dynamical horizons \cite{hay94,ash99, ash00,ash03}, this method offers a huge advantage over other approaches, since these concepts of horizons are quasi-local and independent from global properties such as asymptotic flatness. Due to its very explicit usage of the paths across the horizon, the tunneling picture allows to find generalizations in a very systematic way \cite{sen15,Kant:2009pm}. It also suggests a generalized notion of the Hawking effect \cite{gia20} that applies to all types of horizons, dynamical and static, and draws a connection to the consistency of quantum field theory.

In this article we extend the tunneling method to the case of universal horizons (UH) in Lorentz violating theories of Gravity. In particular, we study Einstein--Aether (EA) gravity \cite{Jacobson:2000xp}, where boost invariance is explicitly broken by the presence of a time-like unit four-vector $U^\mu$, the aether, which propagates a vector and a scalar degree of freedom on top of the usual transverse traceless graviton, when coupled to the Einstein--Hilbert action \cite{Jacobson:2004ts}. As a consequence of this, once the time direction is identified with the integral lines of $U^\m$, the symmetry of the theory is effectively reduced down to foliation preserving diffeomorphisms (FDiff), consisting of the direct product of time reparametrizations and time-dependent spatial diffeomorphisms
\begin{align}\label{eq:FDiff}
    \tau\rightarrow \tau'(\tau),\quad x^i\rightarrow x'^i(\tau,x^i)
\end{align}
where $\tau'(\tau)$ must be a monotonous function\footnote{From now on we use greek indices to denote space-time coordinates -- including time -- and latin indices to denote spatial directions only. We also use a mostly plus convention for the metric signature.}. In other words, once a preferred threading is given by the aether, then the above transformation describes all coordinate changes that respect this particular choice, effectively breaking Lorentz invariance at the local level.

This allows for matter actions coupled to gravity to include higher derivative operators along \emph{spatial directions}, while keeping only two time derivatives, thus avoiding the presence of Ostrogradsky ghost instabilities. As a consequence, matter fields coupled to EA gravity propagate with modified dispersion relations of the generic form
\begin{align}
    \omega^2=k^2+\alpha_4\frac{k^4}{\Lambda^2}
    +\dots +\alpha_{2Z}\frac{k^{2Z}}{\Lambda^{2Z-2}},
\end{align}
which allow for superluminal propagation at high momentum $k> \Lambda$. 

The former statement seems to imply that the usual notion of horizon looses its meaning, since rays of these fields can freely enter and exit the region enclosed by the event horizon of a black hole. However, this naive intuition is broken in the case of known stationary and spherically symmetric black hole solutions in EA Gravity \cite{berg12}. In these, the aether is hypersurface orthogonal, and thus it defines a preferred foliation in co-dimension one hypersurfaces, described by a scalar field $T$
\begin{align}\label{eq:aether_ortho}
    U_\mu = \frac{\partial_\m T}{\sqrt{|\partial_\a T \partial^a T|}},
\end{align}
which has the interpretation of a preferred time direction that every motion has to follow. Incidentally, these are also solutions to the low energy action of Ho\v rava--Lifshitz Gravity \cite{Horava:2009uw,Blas:2009qj}, known as Khronometric Gravity \cite{Blas:2010hb}. The latter propagates only one extra scalar field, in contrast to the vector and scalar propagated by EA gravity. However, in this work we will not discuss the dynamics of gravitational perturbations, and thus all our results can be applied indistinctively to either EA or Ho\v rava--Lifshitz Gravity.

In certain space-times -- which include the black hole solutions of interest --, it might happen that a foliation leaf becomes a constant radius hypersurface. Since the foliation defines a preferred time direction that all world-lines must follow, this implies that nothing can escape the region enclosed by the leaf. Such a hypersurface is thus named \emph{universal horizon} and it is locally characterized by the following two properties \cite{Bhattacharyya:2015gwa}
\begin{align}\label{eq:condition_UH}
    (U\cdot\chi)=0,\quad (a\cdot \chi)\neq 0,
\end{align}
where $\chi^\mu$ is the time-like Killing vector of the metric and $a_\m=U^\a\nabla_\a U_\m$ is the acceleration of the foliation\footnote{We have defined the dot-product between two vectors $X^\mu$ and $Y^\mu$ with respect to the space-time metric to be $(X\cdot Y)=g_{\mu\nu}X^\mu Y^\mu$.}. The usual notion of a horizon as a casual boundary thus gets resurrected in these theories by the existence of the UH. Even those motions travelling at infinite speed will be forever future-trapped in the inner region once they cross the UH. Due to this, one naively expects that UHs must radiate, have an entropy and, in general, replicate all the features that are usually associated to horizons in Lorentz invariant theories.

Indeed, the thermodynamical properties of UHs have been studied in the recent past \cite{he21,mi15}. In \cite{ber13,Ding:2015fyx} the tunneling picture was tentatively applied to UHs. However, despite showing a proof of concept and anticipating the same result we shall find for the Hawking temperature, we feel that in these early works several technical and physical subtleties -- arising from the recently improved understanding on the shape and dynamical role of the preferred foliation -- were not fully spelled out and require a more in detailed discussion. 

In particular, it has been recently shown in \cite{del22} that the two regions bordering the universal horizon must feature a smooth lapse, since otherwise the theory looses its predictability on a fundamental level. This however contradicts the naive expectation that the whole space-time -- exterior and interior -- should be described by the same global foliation or, at the very least, by a discrete set of discontinuous foliations that are, however, globally oriented in the same way. This property manifests itself in the pole structure at the horizon, so that the construction of a well-defined tunneling path is compromised and needs to be revisited. As we shall see in what follows, our renewed analysis of the gravitational tunneling in this framework exposes quite relevant physical lessons concerning the origin and nature of the Hawking radiation in these settings.

In this paper we aim to solve these issues by investigating and formalizing the tunneling picture for UHs in Lorentz-violating theories in a solid way. Our work is organized as follows: First, in Section \ref{sec:relat_tunneling} we summarize and review the standard derivation of Hawking radiation in general relativistic space-times through the tunneling method. Later, in Section \ref{sec:non_re_tunneling} we show how this can be applied to the case of UHs in a straightforward way, by studying arbitrary spherically symmetric solutions and the trajectories of gravitating fields therein, by means of the Hamilton-Jacobi method. Due care will be given to the choice of the privileged time direction in Section \ref{sec:sync_factor}, which influences the temperature of the emitted radiation through the UV behavior of the dispersion relation of the field. Finally, we draw our conclusions in Section \ref{sec:conclusions}.

\section{Relativistic Tunneling}\label{sec:relat_tunneling}
Before we explore the trenches of Lorentz-violating theories, we briefly review the tunneling picture in theories with full invariance under diffeomorphisms. This allows us to introduce the basic concepts that we will later extend to Lorentz-breaking theories. Most importantly, it will helps us clarifying how the tunneling paths are identified, the Hamilton-Jacobi method to describe the tunneling rate, and the consistency criterion for a well-defined Hawking effect. Our review is based on \cite{gia20}, which discusses the relativistic framework generically also for dynamical cases. In the following we restrict ourselves to spherically symmetric and static space-times. However, generalizations are immediate.

\subsection{Gravitational tunneling}

Let us consider an asymptotically flat, static, and spherically symmetric space-time. Its metric can always be written in terms of outgoing (ingoing) rays of light by using Eddington--Finkelstein--Bardeen (EFB) coordinates
\begin{align}\label{efb}
\mbox{d}s^2=-F(r)\mbox{d}\xi^2\pm2 B(r)\mbox{d}\xi\mbox{d}r+r^2\mbox{d}^2S_2,
\end{align}
where d${}^2S_2$ is the line element of the two-dimensional sphere of unit radius, and $\xi=\{u,v\}$ is a placeholder for the light-cone coordinates, $u=t-r^*$ and $v=t+r^*$, with $r^*=\int_r\tfrac{{\rm d}r}{F(r)}$ the tortoise coordinate. Here the functions $F(r)$ and $B(r)$ determine completely the metric. The existence of a horizon is determined by the condition $F(r)=0$, with $F(r)\in\mathcal{C}^2(\mathbb{R})$ at the least. The sign in front of the non-diagonal term is negative (positive) for $\xi\equiv u$ ($\xi\equiv v$).

The horizon, separating trapped and normal regions, acts as a semipermeable surface allowing only penetration from one side. Which paths are classically allowed depend on the nature of the trapped surface. For a black hole, the interior becomes future trapped and only ingoing causal trajectories can cross the horizon. This can be seen from the null-congruences in Schwarzschild space-time, that generate the lightcone. While the ingoing congruence $l^-=-2\partial_r$ is inwards pointing and well defined across the horizon, the outgoing one $l^+=\partial_v+F(r)\partial_r$ changes its direction from outgoing to effectively ingoing, when travelling from the (exterior) normal region into the (interior) trapped region. This is due to the fact that $F(r)$ changes its sign across the horizon. All causal curves in the interior are hence classically destined to stay inside. 

Similarly, causal curves outside the horizon cannot classically have departed from within it. Indeed, they show a singular behavior in their momentum and classical action at the horizon. For example, for the radial momentum $k_r$, one finds through the Hamilton--Jacobi method\footnote{Alternatively, one can use the expansion parameters $\theta^V=h^{ab}\mathcal{L}_Vh_{ab}$ of a hypersurface with metric $h_{ab}$ along a vector field $V^\mu$, that are negative along ingoing and positive along outgoing directions. At the horizon, where one direction gets marginally trapped, we get $\theta=0$. For a Schwarzschild black hole, the outgoing direction along $l^+$ becomes trapped, hence, $\theta^+$ changes sign and vanishes when $F(r)=0$. The classical action $S_0$ can be rewritten using $\theta^+$ such that the spatial momentum $k_r\propto1/\theta^+$, thus inducing a pole in the integral \cite{sen15}.} that $k_r\sim F^{-1}(r)$. A small complexification around the horizon $F(r)\to F(r\pm i\varepsilon)$ can be introduced so to resolve this problem, but this also  leads to  a complex action, given that the imaginary part of the classical action
\be \label{eq:clas_action_C}
\mathcal{S}_0=-\Omega v+\int \frac{2\Omega}{F(r\pm i\varepsilon)}\mbox{d}r,
\ee
is then determined by the Sokhotski--Plemelj theorem\footnote{\label{ft:footnote3} The Sokhotski--Plemelj theorem is a relation for integrals with a small complexification around a pole
\be
\lim_{\varepsilon\to0}\int_a^bdx\ \frac{f(x)}{x\pm i\varepsilon}=\mp i\pi f(0)+\mathcal{P}\left(\int_a^b\frac{f(x)}{x}\right),
\ee
where $\mathcal{P}$ denotes the Cauchy principal value. An alternative path can be taken by complexifying the Hamilton--Jacobi equation by introducing the Feynman $\pm i\varepsilon$-prescription \cite{gia20}.}. Thus, an outgoing trajectory cannot start in the interior of the black hole. The classical one-particle action would become complex along this path, making it classically forbidden \cite{van08}. However, things dramatically change once quantum effects are considered, as  prohibited processes can become quantum-mechanically allowed. In particular, horizon crossing due to gravitational tunneling becomes possible.

Although conceptually different, it is instructive to look first at quantum mechanical tunneling through a finite size barrier. Suppose that we have a bounded one-dimensional $L^1$-integrable classical potential, such that $\lim_{x\rightarrow \pm\infty}V(x)=0$. The probability for a quantum mechanical wave function $\Psi(x)$ with energy $E<{\rm max}\left(V(x)\right)$ to tunnel through it is given by the ratio of the norm of the transmitted wave $\Psi_>$ divided by the norm of the incident wave $\Psi_<$ before hitting the wall
\begin{align}
    |T|^2=\frac{\|\Psi_{>}\|^2}{\|\Psi_{<}\|^2}.
\end{align}

For any shape of the potential, as long as the potential itself is only mildly time-dependent, this can be described by using a WKB-approximation for the wave function
\begin{align}
    \Psi(x)=\Psi_0\exp\left(\tfrac i\hbar\sum_i \mathcal{S}_i(x)\hbar^i\right).
\end{align}

To leading order, and considering the Schr\" odinger equation, the solution for ingoing and outgoing waves is determined by the classical action $\mathcal{S}_0=\pm\int$d$x\sqrt{2m(E-V(x))}$ with $E$ being the energy of the incident wave and $m$ its mass. Thus, one can understand the tunneling process as occurring through a classically forbidden path where the exponent becomes complex -- since inside the potential $E<V(x)$. This has some similarities with the case of gravitational paths, as we see below.

Let us start by noting that gravitational tunneling leads to particle creation by horizons. This has been formulated in \cite{par00,pad02} for the static black hole, and then further refined to a vast variety of examples \cite{van11}. A cartoon picture of the process was already suggested by Hawking \cite{haw75}. Close to but still outside the horizon, a Hawking pair consisting of a positive and a negative energy particle, can be created. The positive energy particle escapes the gravitational well and it is measured by an asymptotic observer, while the negative energy one falls into the black hole (where its existence on-shell is allowed by the spacelike nature of the Killing vector associated to stationarity in time). 

Courtesy of the energy budget provided by the black hole, we can describe this from a different perspective. Instead of a pair creation, we interpret the same process as a single particle coming from the interior and tunneling outwards in a quantum mechanical way. This is possible because a) we can always trade an inward-pointing negative Killing energy particle with a positive energy, outward-pointing one; and b) he notion of energy positivity is linked to the Killing vector, which changes its character at the horizon. 

This idea can be made explicit by considering a massive scalar field $\phi$ in the space-time described by \eqref{efb} and obeying the Klein-Gordon equation
\be\label{kgrel}
\left(\Box-\frac{m^2}{\hbar^2}\right)\phi(x)=0. 
\ee
Assuming that the space-time does not change too rapidly, we can use a WKB ansatz for the field
\be\label{wkb}
\phi=\phi_0e^{\frac i\hbar\sum_n\hbar^n\mathcal{S}_n(x)}=\phi_0 e^{\frac i\hbar \mathcal{S}_0+\mathcal{O}(\hbar^0)},
\ee
with $\mathcal{S}_0(x)$ the classical action. The constant $\phi_0$ is allowed to have a mild coordinate dependence, but it is usually treated as effectively constant. We can then  plug \eqref{wkb} into \eqref{kgrel} to get the Hamilton--Jacobi equation to lowest order\footnote{In principle we could also consider the sub-leading term $\mathcal S_1$. However, this does not contribute to the tunneling probability.} in $\hbar$
\be\label{hje}
\partial_\mu\mathcal{S}_0\partial^\mu\mathcal{S}_0+m^2=0.
\ee

One can formulate a general ansatz for the classical action as $\mathcal{S}_0=\int $d$x^\m\partial_\m \mathcal{S}_0$ with d$x^\m=\sum_K K^\m (K_\n {\rm d}x^\n)/|K|$ where the vectors $K^\m$ determine the observer, or in other words, the vacuum state. They must be such that they span the full chart of space-time coordinates. Generally, some of the $K^\m$ are chosen to reflect the symmetries of the setup, i.e. Killing (static space-times), Kodama (spherically symmetric but dynamical), or dual-null vector (general space-times). In the case of a time-like Killing vector $K^\m\equiv \chi^\m$, this introduces a covariantly conserved energy, or frequency 
\be\label{frq}
\Omega:=-\chi^\m\partial_\m \mathcal{S}_0,
\ee
that defines a preferred notion of observers. The ansatz of $\mathcal{S}_0$, thus, drastically simplifies to
\begin{align}\label{eq:S0_relativ}
    \mathcal{S}_0=-\Omega v + \int k_r \mbox{d}r,
\end{align}
where we have assumed spherical symmetry. Hence, the trajectory is completely determined by the Killing energy and the function $k_r(r)$, denoting the spatial momentum.

\begin{figure}
	\centering
	\includegraphics[width=0.235\textwidth]{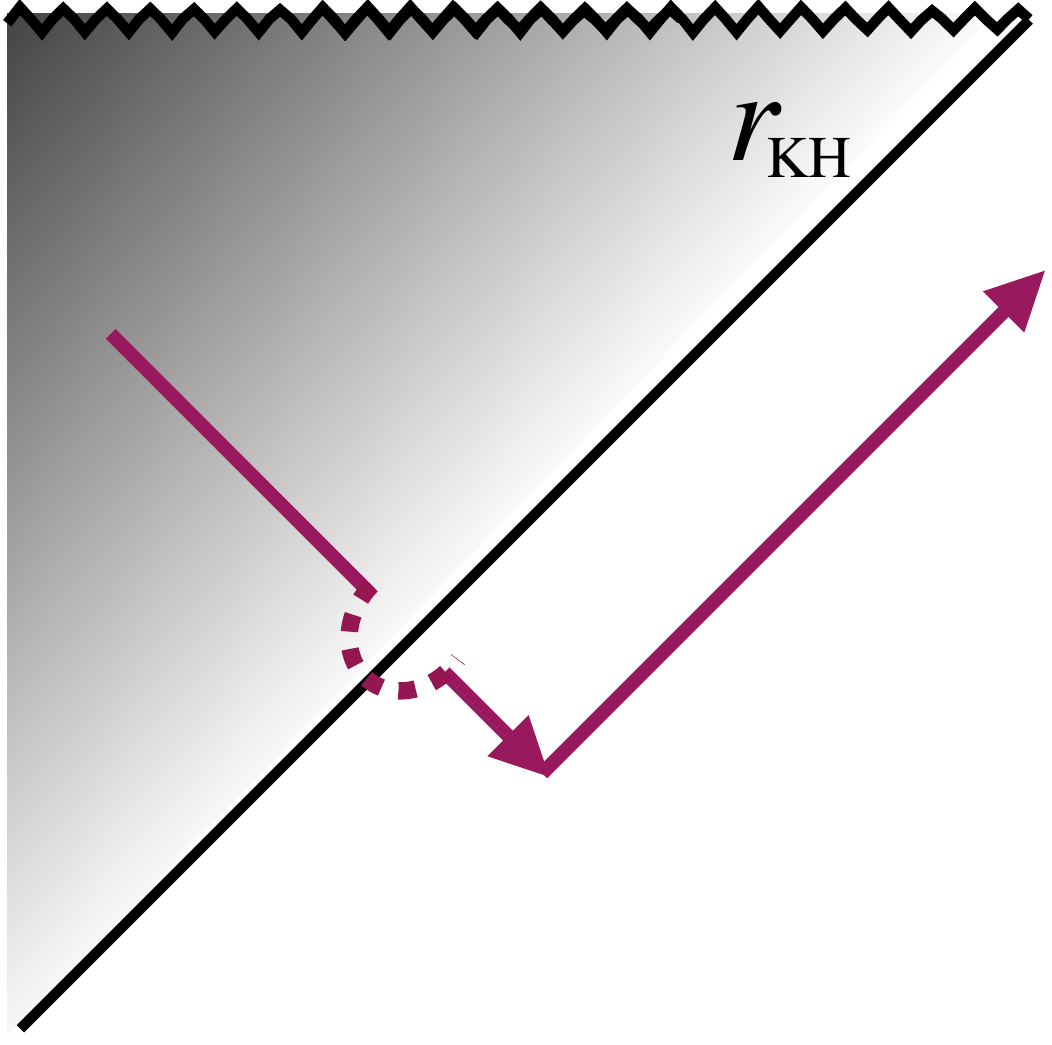}
	\caption{The Killing horizon is displayed as a null-line that separates the normal (unshaded) from the trapped region (grayshaded). The tunneling path shows a positive energy particle that starts in the interior on a past-directed outgoing null path, crosses the horizon on a complex path (dashed), and scatters into a future-directed outgoing null path, once it has crossed the horizon. This is equivalent to the negative energy particle tunneling inwards.}
	\label{eins}
\end{figure}

Coming back to our discussion on the quantum-mechanical tunneling, in a system with gravitational tunneling the probability to reach a classically inaccessible region is given through complex paths, which is reminiscent to the case of the one-dimensional potential barrier. The positive energy particle inside a black hole is now interpreted to take a generically complex path across the horizon determined by $\mathcal{S}_0$, as shown in figure \ref{eins}. 

The particle first travels along a past-directed null-curve to the horizon which becomes an outgoing future-directed null-path after horizon crossing. As we discussed previously in \eqref{eq:clas_action_C}, this happens when $k_r\sim F^{-1}(r)$, which corresponds to the outgoing trajectory. This also establishes a clear link between the existence of tunneling, and hence of particle production, and the presence of a single pole in the classical action\footnote{Higher order poles lead to a purely real action cf. \cite{gelfand,hormander} for details.}.

Finally, we can define the tunneling rate as the ratio between the transmitted fraction and the incident wave \cite{sen15}
\be\label{rate}
\Gamma=\frac{\|\phi_{>}\|^2}{\|\phi_{<}\|^2}\sim e^{-\frac{2}{\hbar}{\rm Im}(\mathcal{S}_0)},
\ee
which vanishes along classical paths ${\rm Im}(\mathcal{S}_0)=0$, but it is finite otherwise. This is connected to thermodynamics by following \cite{har83,gib93}. Comparing the probability for a detector to absorb a particle $P_{\rm abs}$ with its probability to emit one $P_{\rm em}$ at a fixed energy $\Omega$ we get
\be\label{bolli}
\Gamma\equiv\frac{P_{\rm abs}}{P_{\rm em}}=e^{-{\Omega}/{T_{\scaleto{\rm H}{2.5pt}}}},
\ee 
thus finding that the detector is in a thermal equilibrium at (horizon) temperature $T_{\rm H}$. The particle absorbed by the detector coincides with the particle that has crossed the horizon. Thus, whenever ${\rm Im}(\mathcal{S}_0)\propto \Omega$, we can read off a horizon temperature from the tunneling rate. However, the thermodynamics is only well-defined whenever the imaginary part is positive definite \cite{gia20}
\be\label{heb}
\mbox{Im}(\mathcal{S}_0)>0,
\ee
otherwise the process leads to inconsistencies such as the violation of the probabilistic interpretation. In the subsequent analysis we will show the robustness of this criterion even in Lorentz-violating theories, which underlines the resilience of thermodynamic properties and the generality of the analysis in \cite{gia20}.

\subsection{Black holes in the tunneling picture}\label{sec:bh_rel}

As we discussed before, the classical action for an S-wave (spherically symmetric) in EFB coordinates is given by \eqref{eq:S0_relativ}, where all that remains to be done is to obtain the value of $k_r$. This is done by plugging \eqref{eq:S0_relativ} into \eqref{hje}, so arriving at the Hamilton--Jacobi equation in the EFB vacuum
\be
2\Omega k_r-F(r)k^2_r=0.
\ee
Tracing the particle that arrives to the asymptotic region of large radius down to the horizon, we find that its momentum diverges and that the particle thus travels on an approximately light-like trajectory, which allows us to neglect the mass term\footnote{The Hamilton--Jacobi equation is soluble even in the case of keeping the mass term. However, since we will perform later a near horizon approximation, this term will drop out anyways and does not add anything to the analysis but mere complication.} and solve for the radial null momentum $k_r(r)$. One finds two solutions, the first corresponding to the ingoing fields ($k_r=0$), and which is regular across the horizon; and the outgoing momentum \cite{hay09}
\begin{equation}
k_r(r)=\frac{2\Omega}{F(r)},
\end{equation}
which develops the aforementioned pole at the position of the horizon, thus requiring to be analytically continued in $r$ through $F(r)=0$. In our case we choose\footnote{The definition for a quantum process leading to a consistent contour integral, as well as thermodynamics, was derived in \cite{gia20} and is given by $\mbox{Im}(\mathcal{S}_0)>0$. This influences our choice of analytic continuation. Note, that once a physical process is determined, the result does not depend on the choice of the analytic continuation. This can be seen by its independence of the $i\varepsilon$-prescription.} $F(r)\rightarrow F(r-i\varepsilon)$, with $\varepsilon \ll 1$. Alternatively, one can add a Feynman $i\varepsilon$-prescription in the Hamilton--Jacobi equation and solve directly for the imaginary part. As a result, $k_r(r)$ will contribute to Im$(\int$d$r k_r(r))$ such that after some manipulations (cf. \cite{gia20} for details)
\be\label{ims0rel}
\mbox{Im}(\mathcal{S}_0)=\lim_{\varepsilon\to0}\mbox{Im}\left(\int_{r_1}^{r_2}\frac{\Omega\mbox{d}r}{ \kappa_{\textsc{kh}}
\left(r-r_{\textsc{kh}}-\frac{i\varepsilon}{\kappa_{\textsc{kh}}
}
\right)}\right)=\frac{\pi\Omega}{\kappa_{\textsc{kh}}},
\ee
where we performed a near-horizon expansion $F(r)\cong\kappa_{\scaleto{\rm KH}{3pt}}(r-r_{\scaleto{\rm KH}{3pt}})+\mathcal{O}\left((r-r_{\scaleto{\rm KH}{3pt}})^2\right)$, and identified the surface gravity of the horizon as $\kappa_{\scaleto{\rm KH}{3pt}}=\tfrac{\partial F}{\partial r}|_{r=r_{\scaleto{\rm KH}{2.5pt}}}$. This surface gravity can be shown to fulfill the condition $l^\mu\nabla_\mu l^\nu=2\kappa_{\scaleto{\rm KH}{3pt}}l^\nu$ that measures the inaffinity properties of null-geodesics \cite{hay09}.

Furthermore, we implemented the crossing path of the positive energy mode from the interior starting at $r_1<r_{\scaleto{\rm KH}{3pt}}$ and tunneling to the exterior region to $r_2>r_{\scaleto{\rm KH}{3pt}}$ \cite{par00}. Comparing the tunneling rate \eqref{rate} with the Boltzmann distribution \eqref{bolli} we find
\be\label{reltemp}
\Gamma=e^{-{2\pi\Omega}/{\kappa_{\textsc{kh}}}}= e^{-{E}/{T_{\textsc{h}}}},
\ee
yielding the standard result for the horizon temperature $T_H=\frac{\kappa_{\textsc{kh}}}{2\pi}$.

\section{Non-relativistic tunneling} \label{sec:non_re_tunneling}
After summarizing the tunneling approach in relativistic theories, let us now turn to the main topic of this work. In the following, we will extend the tunneling picture to the case of space-times endowed with a UH in Lorentz violating theories. In particular, we will focus on EA gravity, which for our purposes here also includes the low energy limit of Ho\v rava--Lifshitz Gravity. In what follows, we investigate tunneling across UHs, and hereinafter verify the principles formulated in \cite{gia20}, and revisited in the previous section, for Lorentz breaking theories. Their persistence even in absence of Lorentz symmetry fortifies the universality of horizon thermodynamics. 

\subsection{Einstein--Aether Gravity}
We consider matter actions coupled to EA gravity, with action
\begin{align}\label{eq:EA-action}
    I_{\rm EA}=\frac{1}{16\pi G}\int \mbox{d}^4x \sqrt{-g} \left(R-{\cal L}_{\rm U} + \lambda(U_\m U^\m+1)\right),
\end{align}
where $R$ is the Ricci scalar, $\lambda$ is a Lagrange multiplier implementing a unit norm condition for the aether field $U^\mu$, and ${\cal L}_{\rm U}=K^{\a\b}_{\m\n}\nabla_\a U^\m \nabla_\b U^\n$, with $K^{\a\b}_{\m\n}=c_1 g^{\a\b}g_{\m\n}+c_2 \delta^\a_\m \delta^\b_\n +c_3 \delta^\a_\n \delta^\b_\m + c_4 U^\a U^\b g_{\m\n}$ and couplings $c_i\in\mathbb{R}$. 

This action is invariant under the FDiff transformations \eqref{eq:FDiff}, which then dictate the form of the matter action. Hereinafter we take the simplest case of a Lifshitz scalar field\footnote{Here we set a possible mass term to vanish. As in the relativistic tunneling, its presence is irrelevant and thus we can neglect it without loss of generality.}
\begin{align}\label{eq:lifshitz}
I_{\rm m}=-\frac12\int {\rm d}^4x \sqrt{-g}\left[\partial^\m\phi\partial_\m\phi+\sum_{z=2}^Z\frac{\alpha_{2z}}{\Lambda^{2z-2}}\phi(-\Delta)^z\phi\right],
\end{align}
whose corresponding equation of motion is thus
\begin{align}\label{eq:lifhistz_eom}
    \square \phi -\sum_{z=2}^Z\frac{\alpha_{2z}}{\Lambda^{2z-2}}(-\Delta)^z\phi=0.
\end{align}

Here, all the $\alpha_{2z}$ are dimensionless, and we choose to normalize $\alpha_{2Z}=1$. The scale $\Lambda$ sets the momentum scale at which Lorentz violations become relevant. In 3+1 dimensions, usually $Z=3$ is chosen so to enforce power-counting renormalizability of the gravitational action in the case of Ho\v rava--Lifschitz (HL) gravity \cite{Pospelov:2010mp,Anselmi:2007ri,Barvinsky:2015kil}. The d'Alembert as well as the spatial Laplace operators are given by $\square=g^{\m\n}\nabla_\m\nabla_\n$ and $\Delta=\gamma^{\m\n}\nabla_\m\nabla_\n$, with $\gamma_{\m\n}=g_{\m\n}-U_\m U_\n$ the metric induced on the leafs orthogonal to the aether. We further restrain ourselves to space-times with UHs, for which it is a necessary condition that the aether is hypersurface orthogonal, taking the form \eqref{eq:aether_ortho}. Examples of vacuum solutions of this kind can be found analytically \cite{berg12} and numerically \cite{Barausse:2011pu} for generic regions of the parameter space of EA gravity. All of them equally correspond to vacuum solutions of Khronometric gravity \cite{Blas:2010hb}. Finally, we will also require the existence of a time-like Killing vector $\chi^\m$ -- i.e.~both metric and aether are supposed to be Lie dragged by this field --, from which a notion of a conserved Killing energy for a particle of four-momentum $k^\mu$, namely $\Omega=-(\chi\cdot k)$,  can be derived.

Note that the equation of motion \eqref{eq:lifhistz_eom} generically contains higher spatial and time derivatives, since $\gamma^{00}$ and $\gamma^{0i}$ are generically non-vanishing. This fact is quite problematic, because it leads to Ostrogradsky ghosts, signaling a classical runaway instability and a loss of unitarity in the quantum theory. Only when the time direction is identified with the integral lines of $U^\m$, the equation of motion remains second order in time derivatives. This selects a preferred time direction that every motion must follow within this space-time, and thus a universal preferred frame, up to FDiff transformations of course. In this frame, the scalar field $\phi$ inherits a modified dispersion relation which can be read from \eqref{eq:lifhistz_eom}. 

Given a particle four-momentum $k_\mu\phi=-i(\partial_\mu\phi)$, it can always be decomposed in the aether frame as
\begin{align}\label{eq:k_vector}
    k^\m=\omega U^\m+k_\rho S^\m,
\end{align}
where $S^\m$ is the space-like unit vector orthogonal to $U^\m$.
Here $\omega$ and $k_\rho$ are the energy and spatial momentum in the aether frame
\begin{align}
    \omega=-(U\cdot k),\quad k_\rho =(S\cdot k).
\end{align}
Hence, in the preferred frame \eqref{eq:lifhistz_eom} implies the dispersion relation
\begin{align}
   \omega^2=k_\rho^2+\sum_{z=1}^Z \alpha_{2z}\frac{k_\rho^{2z}}{\Lambda^{2z-2}}.
\end{align}
Note however that these aether frame quantities are \emph{not constant} along the motion. As said before, it is the Killing energy to be conserved along free motions under the sole influence of metric and aether. Given the above definitions we decompose it as
\begin{align}
  \Omega=  (\chi\cdot k)=\omega(U\cdot \chi)-k_\rho (S\cdot \chi),
  \label{eq:decOm}
\end{align}

\subsection{Spherically symmetric black holes}

In the following we consider arbitrary spherically symmetric and static solutions to the equations of motion obtained from \eqref{eq:EA-action}, written in the form \eqref{efb} with $\xi\equiv v$
\begin{equation}
\mbox{d}s^2=-F(r)\mbox{d}v^2+2B(r)\mbox{d}v\mbox{d}r+r^2\mbox{d}^2S_2,
\end{equation}
and endowed with an aether $U^\m$ of the form~\cite{berg12}
\begin{equation}
U_\mu \mbox{d}x^\mu=-\frac{1+F(r)A^2(r)}{2 A(r)}\mbox{d}v+B(r)A(r)\mbox{d}r,\quad S_\m \mbox{d}x^\mu=\frac{1-F(r)A^2(r)}{2 A(r)} \mbox{d}v+B(r)A(r)\mbox{d}r,
\end{equation}
satisfying $U_\m U^\m=-1$. As said above, $S^\m$ is the vector orthogonal to $U^\m$, which satisfies $S_\m S^\m=1$, and it is chosen to be inwards pointing. The functions $F(r)$, $B(r)$, and $A(r)$ must be determined case by case by solving explicitly the equations of motion derived from \eqref{eq:EA-action}. Hereinafter we keep the functions arbitrary\footnote{In spherically symmetric setups, a generic vector contains two independent functions. The normalization condition of the aether fixes one of them, and as such only one variable $A(r)$ is left. The specific form of $U$ is then chosen for computational convenience.}, albeit demanding the existence of a Killing as well as a universal horizon. The former is localized by the condition $F(r_{\scaleto{\rm KH}{3pt}})=0$, while the latter is given by
\begin{align}
\left.(U\cdot\chi)\right|_{\scaleto{\rm UH}{3pt}}=-\frac{1+F(r_{\scaleto{\rm UH}{3pt}})A^2(r_{\scaleto{\rm UH}{3pt}})}{2 A(r_{\scaleto{\rm UH}{3pt}})}=0,
\end{align}
with $\chi^\m=(1,0,0,0)$ the time-like Killing vector of the metric. Incidentally, note that  the lapse $N$ of the foliation described by the surfaces orthogonal to the aether field (cf. below), is $N=-(U\cdot\chi)$ and so it changes sign crossing the UH \cite{del22}. Finally, we also demand asymptotic flatness, which implies that $U^\m$ becomes parallel to $\chi^\m$ at large radius, thus fixing the asymptotics of $F(r)$ and $A(r)$ to $F(r)=A(r)=1$ for $r\rightarrow \infty$Note, that there exist also black hole solutions with maximally symmetric asymptotics \cite{b14}.

Since the aether defines a physical foliation, it is convenient to change coordinates into the preferred system, that allows us to align the notion of time with the evolution direction of the aether, such that the spatial vector will always be tangent to the foliation. The following transformation allows to change into this system, given by $\{\tau,\rho,\vartheta,\varphi\}$ \cite{cro14}
\begin{equation}\label{trafo}
\tau=v+\int\frac{U_r}{U_v}\mbox{d}r\quad\mbox{and}\quad\rho=\int\frac{S_r}{S_v}\mbox{d}r+v.
\end{equation}

There is an important point to take into account here, since the integral defining $\tau$ takes the form
\begin{equation}
\int\frac{U_r}{U_v}\mbox{d}r=\int \frac{-2B(r)A^2(r)}{1+F(r)A^2(r)}\mbox{d}r,
\end{equation}
it actually diverges at the UH and in its proximity behaves as
\begin{align}
    \int\frac{U_r}{U_v}\mbox{d}r=\frac{B(r_{\scaleto{\rm UH}{3pt}})A(r_{\scaleto{\rm UH}{3pt}}) }{\left. \partial_r N\right|_{\scaleto{\rm UH}{3pt}} } \ln\left|r-r_{\scaleto{\rm UH}{3pt}}\right|+{\cal O}(r-r_{\scaleto{\rm UH}{3pt}}),
\end{align}
where we have expanded the lapse as $N=\left. \partial_r N\right|_{\scaleto{\rm UH}{3pt}} (r-r_{\scaleto{\rm UH}{3pt}})$, and taken into account that $A(r)$ and $B(r)$ are regular everywhere. In these coordinates, the position of the UH thus corresponds to $\tau\rightarrow \infty$, with the foliation exhibiting an infinite accumulation there. This matches the behavior of the light-cone coordinate $u$ in standard approaches to Hawking radiation (cf. \cite{Jacobson:2003vx} and references within), and hence serves as a smoking gun to expect a similar property from the UH. Moreover, we can also note that the logarithm requires an analytic continuation through $r=r_{\scaleto{\rm UH}{3pt}}$. This will introduce an imaginary part that will eventually resurface in the solution to the Hamilton--Jacobi equation for the field in \eqref{eq:lifshitz}, in similar manner to the relativistic case discussed in section \ref{sec:bh_rel}.

In foliation adapted coordinates the metric takes the ADM form
\begin{equation}\label{metricpc}
\mbox{d}s^2=-(N^2-N^i   N_i)\mbox{d}\tau^2+2N_\rho \mbox{d}\tau\mbox{d}\rho+\gamma_{ij}\mbox{d}x^i \mbox{d}x^j,
\end{equation}
where the Latin indices now run over the spatial directions $\{\rho, \vartheta,\varphi\}$ only. The lapse $N$, shift vector $N^i$, and spatial metric $\gamma_{ij}$, are given by
\begin{align}\label{eq:lapse}
    &N=\frac{1+F(r)A^2(r)}{2 A(r)},\\
    &N^i=0,\\
    &\gamma_{ij}\mbox{d}x^i \mbox{d}x^j=\frac{(1-F(r)A^2(r))^2}{4 A^2(r)}\mbox{d}\rho^2+r^2\mbox{d}^2S_2,
\end{align}
where $r$ must be understood as $r(\tau,\rho)$ at all times. Notice that the conditions for the existence of the UH \eqref{eq:condition_UH} translate in this language to $N=0$ and $\partial_r N\neq0$, which again points towards a full analogy with the relativistic case. This criterion is fully consistent with the covariant criterion given in \cite{di09}.

We thus proceed as in the relativistic case, by constructing a solution for the Lifshitz scalar field $\phi$ by means of the WKB approximation, obtaining the corresponding  Hamilton--Jacobi equation. To this aim, we need the inverse metric
\begin{equation}\label{eq:inverse_metric}
g^{-1}=\mbox{diag}\left(-\frac{1}{N^2},\gamma^{\rho\rho},\frac{1}{r^2},\frac{1}{r^2\sin^2(\vartheta)}\right).
\end{equation}
All divergences when approaching the UH are encoded in the behavior of the lapse, so mere observation is enough to spot that at the UH, only $g^{00}$ blows up, while the other components remain finite. Note also that close to the Killing horizon all coefficients are regular. This will be important in stressing the difference between the Killing and the universal horizon in the setting at hand.

\subsection{WKB approximation and Hamilton--Jacobi equation}
In order to derive the Hamilton--Jacobi equation we first need to formulate an ansatz for our field through the WKB approximation \eqref{wkb}. 
 Again, in the semiclassical limit, the dominant contribution is given by the classical action $\mathcal{S}_0$.  Using the ansatz\footnote{From now on, we work in natural units such that although formally $\hbar$ plays the role of a smallness parameter that is later identified with the physical $\hbar$, we will set it to one.}
\begin{equation}\label{s0dec}
\mathcal{S}_0=\int\mbox{d}x^\mu\partial_\mu\mathcal{S}_0=-\int \omega\mbox{d}\tau+\int k_\rho\mbox{d}\rho,
\end{equation}
where we have assumed spherically symmetric waves. Using it, we can obtain the corresponding leading order of the Hamilton--Jacobi equation by inserting the WKB approximated field in \eqref{eq:lifhistz_eom} to get
\begin{equation}\label{hamjac}
-\frac{\omega^2}{N^2}+\gamma^{\rho\rho}k_\rho^2+\frac{(\gamma^{\rho\rho})^Z}{\Lambda^{2Z-2}}\left(k_\rho^{2Z}+G(k_\rho,\nabla k_\rho;\Lambda)\right)=0.
\end{equation}
The operator $G(k_\rho,\nabla k_\rho;\Lambda)$ contains subleading terms that depend on a combination of $k_\rho$ and derivatives thereof, but which are suppressed at the UH. Since the leading contribution is given by the highest power, we singled this term out.

\begin{figure}[]
	\centering
	\includegraphics[width=0.4\textwidth]{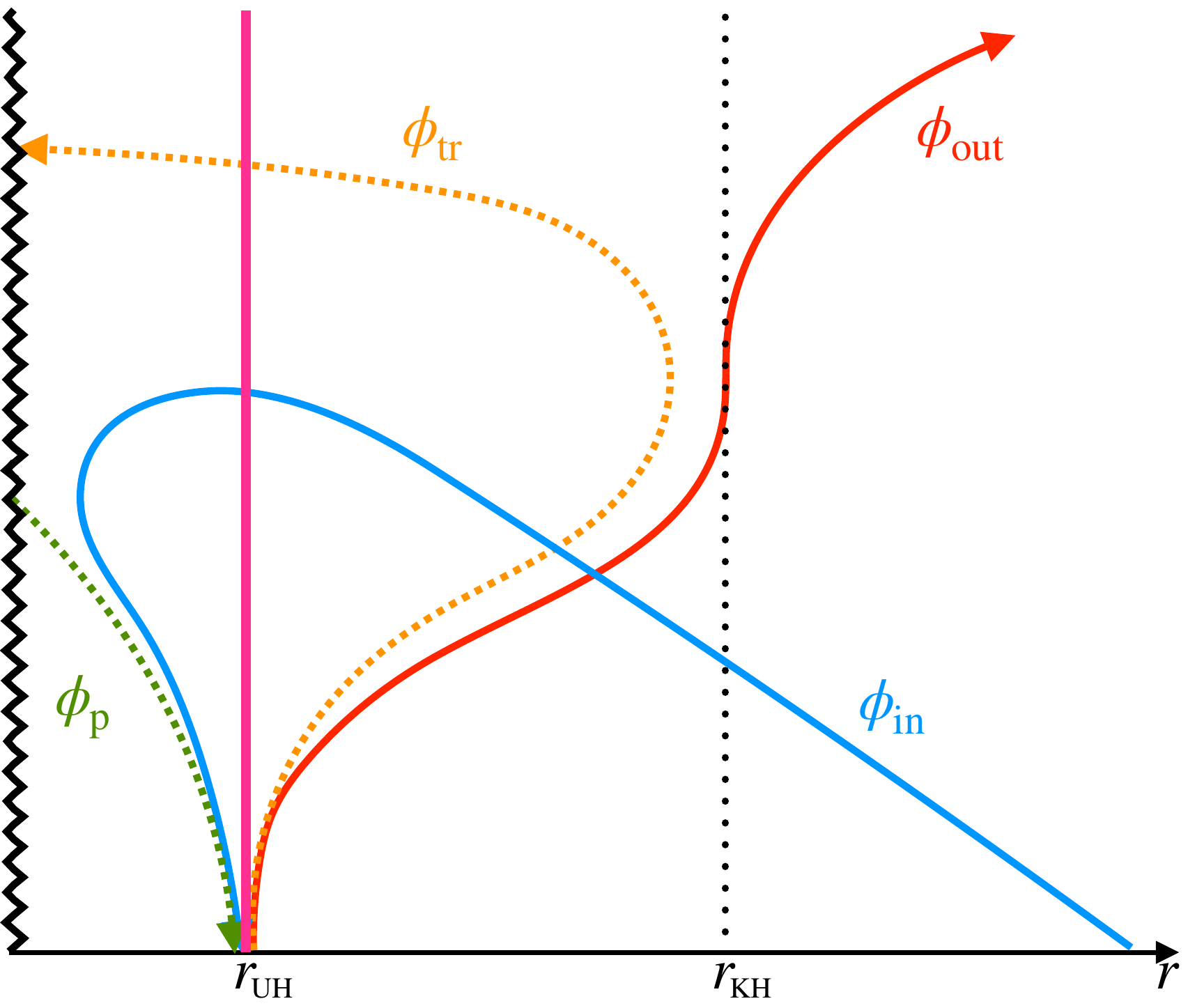}
	\caption{Structure of the modes. Close to the UH and outside of it we have four modes -- ingoing $\phi_{\rm in}$, outgoing $\phi_{\rm out}$, and the two tails of the trapped one $\phi_{\rm tr}$. However, the latter is indistinguishable from the former in the region $r\sim r_{\rm UH}$. Note that the mode $\phi_{\rm p}$ is the Hawking partner to the red mode. The UH is denoted by the purple line, while the Killing horizon is showed as dotted line. The zig-zag line on the left depicts the singularity.} \label{pmodes}
\end{figure}

To understand better the behavior of the quantum fields across the horizon, we investigate the structure of \eqref{hamjac} by following the discussion in \cite{mi15,he21}. We distinguish two regimes -- the \emph{soft regime}, corresponding to $\Lambda\to 0$, and the \emph{hard regime}, where we take $\Lambda\to \infty$ instead. The former displays solutions that cross the universal horizon with finite momentum, while in the latter they have been red-shifted by climbing up the gravitational well, such that their momentum diverges when traced back to the universal horizon, lingering there eternally. As such, these hard modes show a non-analytic behavior in $(r-r_{\rm UH})$, in contrast to those found in the soft regime\footnote{Note that the \emph{softness} is related to the character of the modes when crossing the UH, and nothing prevents soft modes from carrying large momentum.}. A detailed discussion of the behavior of all solutions can be found in \cite{mi15,he21}, and it is summarized in Figure \ref{pmodes}.

At any moment, outside the universal horizon but still nearby, we have four solutions. The ingoing blue mode $\phi_{\rm in}$ (soft), the outgoing red mode $\phi_{\rm out}$ (hard), and the \emph{trapped} orange mode $\phi_{\rm tr}$, which departs the UH as hard, but eventually turns back, crossing it as a soft mode. Although this mode is clearly different form the other two, it mimics their behaviors when we zoom close to the UH. In order to single out the orange mode, we would need to trace it across the region enclosed between the two horizons. For our purposes here, it is thus enough to talk about soft and hard modes, since the trapped one degenerates locally with the ingoing and outgoing rays. 

In the interior of the UH we find a similar structure, although with the roles of $\phi_{\rm tr}$ and $\phi_{\rm in}$ exchanged. While the former remains soft, the latter becomes hard when approaching the UH from its interior. In this region we also find a purely hard mode $\phi_{\rm p}$, corresponding to the CP reversed --- and thus also T reversed, provided that CPT is conserved in our framework --- partner of $\phi_{\rm out}$. This mode is equally diverging as the latter in the UH, but travels from the singularity towards the horizon\footnote{Note however that the direction of the evolution of the rays is tied to the notion of time advance as measured by an observer. The description depicted here corresponds to an observer sitting \emph{outside} the UH. A different observer sitting on its interior would instead measure a reversed time flow -- because the sign of their lapse is flipped, cf. \cite{del22} -- and would assign a reversed evolution to the modes.}.

Once this is settled, and in order to understand the contribution of all these modes to the tunnelling probability, we find the ``soft" and ``hard" limits of the Hamilton--Jacobi equation to be
\begin{eqnarray}
-\frac{\omega^2}{N^2}+\gamma^{\rho\rho}k_\rho^2+m^2+\mathcal{O}\left(\frac{1}{\Lambda}\right)&=&0,\label{shje}\qquad\mbox{Soft}\\
\frac{1}{\Lambda^{2Z-2}}\left(k_\rho^{2Z}+G(k_\rho,\nabla k_\rho;\Lambda)+\mathcal{O}(\Lambda)\right)&=&0. \qquad\mbox{Hard}\label{hhje}
\end{eqnarray}

As previously mentioned however, and differently from the Killing energy $\Omega$, neither $\omega$ nor $k_\rho$ are constants of motion, which obstructs the efforts of solving the previous equations explicitly. However, we can make use of the relation Eq.~\eqref{eq:decOm} among these quantities to write
\begin{equation}\label{kleinomega}
\omega=-\frac{k_\rho (S\cdot\chi)+\Omega}{(U\cdot\chi)}.
\end{equation}

We also need to solve $k_\rho$ in terms of $\Omega$ and its associated spatial momentum $k_r=(k\cdot \zeta)$, where $\zeta^\m$ is the vector orthogonal to $\chi^\m$ and satisfying $\zeta^2=-\chi^2$, taken inwards pointing. Contracting now \eqref{eq:k_vector} with $\zeta^\mu$ and solving for $k_\rho$ we get
\begin{align}\label{krho}
    k_\rho=-\Theta \left(\Omega+\frac{(U\cdot \chi)}{(U\cdot \zeta)}k_r\right),
\end{align}
where we have defined the position dependent angle to be
\begin{align}
    \Theta=\left[(S\cdot\chi)-\frac{(S\cdot\zeta)(U\cdot\chi)}{(U\cdot \zeta)}\right]^{-1}.
\end{align}
Asymptotically $\Theta$ vanishes, while close to the UH, we find instead
\begin{align}\label{eq:theta_limit}
    \left.\Theta\right|_{\rm UH}=(S\cdot\chi)^{-1}.
\end{align}
By substituting these quantities into the Hamilton--Jacobi equation, we have all the required tools to study the contribution of all the different modes to the tunneling probabilities.

\subsubsection{Soft Hamilton--Jacobi equation}
Let us start by considering the soft Hamilton--Jacobi equation \eqref{shje} with the substitutions \eqref{kleinomega} and \eqref{krho}. This introduces additional dependencies on the lapse function, as we can see by noting that $(U\cdot\chi)=-N$. Additionally, we expect these modes to cross the universal horizon with finite momentum. Taking this into account and working close to the universal horizon, so that $N\sim \left. \partial_r N\right|_{\scaleto{\rm UH}{3pt}}(r-r_{\scaleto{\rm UH}{3pt}})$, the leading term in \eqref{shje} becomes
\begin{align}
    \frac{\left(\Omega+k_\rho (S\cdot\chi)\right)^2}{N^4}+{\cal O}\left(\frac{1}{N^2}\right)=0,
\end{align}
with solution
\begin{align}
k_\rho=-\frac{\Omega}{(S\cdot\chi)}.
\end{align}
In terms of $k_r$ this turns out to be
\begin{align}
    k_r^{\rm s}=\Omega\left[1-\Theta (S\cdot \chi)\right]\frac{(U\cdot \zeta)}{(U\cdot \chi)}=-\Omega\left[1-\Theta (S\cdot \chi)\right]\frac{(U\cdot \zeta)}{N}.
\end{align}
Here we have indicated the soft (and later the hard) modes with the superscript `s' (and `h' respectively).

What is left is to evaluate $\mathcal{S}_0$ to obtain its imaginary part, if any. In order to do this, we integrate the solution above through a contour integral connecting the exterior and interior of the universal horizon. Since the soft modes are regular along the UH, they have support along $r=r_{\scaleto{\rm UH}{3pt}}$ and the integration is straightforward
\begin{align}
{\rm Im}\left(\mathcal{S}^{\rm s}_0\right)&={\rm Im}\left(\oint_c k^{\rm s}_{r}(r)\mbox{d}r\right)=-\frac{\Omega \left[1-\left.\Theta\right|_{\scaleto{\rm UH}{3pt}} (S\cdot \chi)_{\scaleto{\rm UH}{3pt}}\right](U\cdot \zeta)_{\scaleto{\rm UH}{3pt}}}{\left. \partial_r N\right|_{\scaleto{\rm UH}{3pt}}}{\rm Im}\left(\oint_c\frac{\mbox{d}r}{r-r_{\scaleto{\rm UH}{3pt}}}\right),
\end{align}
where we have used $(U\cdot \chi)=-N$. Using again the Sokhotski--Plemelj theorem (cf. footnote \ref{ft:footnote3}) to resolve the pole in the lapse, this evaluates to
\begin{align}
   {\rm Im}\left(\mathcal{S}^{\rm s}_0\right) =\frac{\pi \Omega \left[1-\left.\Theta\right|_{\scaleto{\rm UH}{3pt}} (S\cdot \chi)_{\scaleto{\rm UH}{3pt}}\right](U\cdot \zeta)_{\scaleto{\rm UH}{3pt}}}{\left. \partial_r N\right|_{\scaleto{\rm UH}{3pt}}}=0,
\end{align}
which vanishes due to \eqref{eq:theta_limit}. This implies that the modes following these trajectories do not escape the gravitational well. This is in agreement with their description as purely incoming, and thus classically allowed, through the UH.

\subsubsection{Hard Hamilton--Jacobi equation}
Now we take the version \eqref{hhje} of the Hamilton--Jacobi equation and proceed similarly to the previous case. Note that although it does not look dynamical, it yields a dynamical equation for the free-falling observer once the relation \eqref{krho} is imposed. The obvious solution to \eqref{hhje} is $k_\rho=0$, which in turns implies
\begin{equation}\label{hardmodes}
    k_\rho=0\quad \longrightarrow\quad  k^{\rm h}_r=- \Omega\times\frac{(U\cdot \zeta)}{(U\cdot\chi)}.
\end{equation}

As before, we again integrate $k^{\rm h}_{r}$ close to the universal horizon through a contour integral (cf. footnote \ref{ft:footnote3}). However, this time we find that the hard modes do not have support through the UH, so the tunneling path has to be made out of combining an interior hard mode -- either $\phi_{\rm p}$ or $\phi_{\rm in}$ -- and an exterior one -- $\phi_{\rm out}$ or $\phi_{\rm tr}$. In practice, however, this simply corresponds to analytically continuing one of the modes through the UH, while ensuring that $\partial_r N$ is continuous -- cf. \cite{del22}. Moreover, by combining the modes in this way, we find a precise reproduction of Hawking's cartoon picture of gravitational tunneling.

Performing the integral, this time we find a non-trivial contribution to the imaginary part 
\begin{equation}\label{eq:result_hard}
{\rm Im}\left(\mathcal{S}^{\rm h}_0\right)={\rm Im}\left(\oint k^{\rm h}_{r}(r)\mbox{d}r\right)=\Omega \times \frac{ (U\cdot \zeta)_{\scaleto{\rm UH}{3pt}}}{\left. \partial_r N\right|_{\scaleto{\rm UH}{3pt}}}{\rm Im}\left(\oint\frac{\mbox{d}r}{r-r_{\scaleto{\rm UH}{3pt}}}\right)=\pi \Omega\times \frac{(U\cdot \zeta)_{\scaleto{\rm UH}{3pt}}}{\left. \partial_r N\right|_{\scaleto{\rm UH}{3pt}}}=\frac{\pi \Omega  \sqrt{-F(r_{\scaleto{\rm UH}{3pt}})}}{\left. \partial_r N\right|_{\scaleto{\rm UH}{3pt}}},
\end{equation}
where we have used \eqref{eq:theta_limit} and evaluated $(U\cdot\zeta)_{\scaleto{\rm UH}{3pt}}= \sqrt{-F(r_{\scaleto{\rm UH}{3pt}})}$.

In contrast to the ingoing mode, we obtain here instead a non-vanishing imaginary contribution to $\mathcal{S}^{\rm h}_0$. This highlights the fact that the outgoing mode corresponds to trajectories whose momentum diverges at the universal horizon, developing a pole which demands an analytical continuation. Classically, these are unable to cross the surface at $r=r_{\scaleto{\rm UH}{3pt}}$, but quantum mechanical tunneling makes this possible.

Finally, as in the relativistic case, we can obtain the temperature of the radiation composed by these modes through a comparison with the Boltzmann distribution
\begin{equation}
\Gamma=\frac{\|\phi_>\|}{\|\phi_<\|}\propto e^{-2{\rm Im}(\mathcal{S}_0)}\equiv e^{-\Omega/ {T_{\scaleto{\rm H}{2.5pt}}}}
\end{equation}
so that
\begin{equation}\label{eq:temp}
T_{\rm H}=\frac{\left. \partial_r N\right|_{\scaleto{\rm UH}{3pt}}}{2\pi\sqrt{-F(r_{\scaleto{\rm UH}{3pt}})}}.
\end{equation}

Note that our result for 
${\rm Im}(\mathcal{S}_0)$ also fulfills the property \eqref{heb} which shows that this setup admits well-defined thermodynamics. Note in particular that in the case of UHs discussed here, the existence of an analytic continuation of $N^{-1}$ through $r=r_{\scaleto{\rm UH}{3pt}}$ requires $\left.\partial_r N\right|_{\scaleto{\rm UH}{3pt}}$ to be continuous across the UH, thus connecting the consistency of the tunneling approach with the results of \cite{del22}.

Finally, we just need to evaluate the value of $\left.\partial_r N\right|_{\scaleto{\rm UH}{3pt}}$. In order to do that, we construct the acceleration of the aether $a^\m=U^\n\nabla_\n U^\m$, and contract it with the Killing vector, getting
\begin{align}
    (\chi\cdot a)=U^\n \partial_\n (U\cdot\chi)=-U^r \partial_r N,
\end{align}
where we have used that the metric and aether are static and that $\chi^\mu$ is a Killing vector. Evaluating it close to the UH we find
\begin{align}
    - \frac{\left.\partial_r N\right|_{\scaleto{\rm UH}{3pt}}}{\sqrt{-F(r_{\scaleto{\rm UH}{3pt}})} } = (\chi\cdot a)_{\scaleto{\rm UH}{3pt}} = -2 \kappa_{\scaleto{\rm UH}{3pt}},
\end{align}
where $\kappa_{\scaleto{\rm UH}{3pt}}$ is the surface gravity of the UH as derived in \cite{cro14}, controlling the peeling of constant khronon hypersurfaces away from the UH. This is in analogy to the standard surface gravity of a Killing horizon, which controls the peeling of null rays away from it. Substituting this onto \eqref{eq:temp} we finally arrive at the horizon temperature
\begin{align}
    T_{\rm H}=\frac{\kappa_{\scaleto{\rm UH}{3pt}}}{\pi}.
\end{align}
This result agrees with that obtained in \cite{del22} by other methods.

\subsubsection{Radiation from the Killing horizon?}\label{sec:killing_radiation}
The analysis performed in this section hints towards the conclusion that the UH controls the thermodynamic properties of the solution, rather than the Killing horizon. To confirm this, we can observe what happens at the latter. In this region of space-time, neither the elements of the inverse metric \eqref{eq:inverse_metric}, nor the relations \eqref{kleinomega} and \eqref{krho} between aether frame and EFB frame quantities diverge at the Killing horizon. Due to this, the momentum of the modes will also be regular when crossing the Killing horizon, signaling that it is not a causal barrier anymore and can be exited in a finite -- perhaps long -- time. Thus, the integral $\oint k_r(r)$d$r$ along a path crossing the Killing horizon will be strictly real for all the modes, no matter their character. Since no mode develops a pole at the Killing horizon, there is no need for an analytic continuation. This shows that the UH is the sole responsible for the thermodynamical properties of the system, being also the only true causal boundary within the spacetime. The role of the Killing horizon must be, at most, to introduce a graybody factor that might distort the shape of the distribution measured by an observer sitting at large radius, compared to the one emitted close to the universal horizon. However, this analysis is beyond the scope of this work.

\section{Time reparametrization invariance and UV sensitivity of the temperature}\label{sec:sync_factor}

In the previous section we obtained the temperature of the radiation emitted by the UH via the Hamilton--Jacobi method and the ansatz \eqref{s0dec}. The reader might notice that in writing this ansatz we chose a particular foliation time $\tau$, which identifies the lapse function of the foliation by $N=-(U\cdot\chi)$. However, this construction is far from unique. Indeed, as discussed in the introduction, EA configurations are invariant under time reparametrizations
\begin{align}\label{eq:time_rep}
    \tau\rightarrow \hat\tau(\tau),
\end{align}
which preserve the form of $U^\mu$, but not the choice of the lapse function. Under a transformation \eqref{eq:time_rep} the lapse function transforms as
\begin{align}\label{eq:lapse_relation}
    \hat{N}=\frac{{\rm d}\tau}{{\rm d}\hat \tau} N,
\end{align}
which has a strong effect on the value of the temperature \eqref{eq:temp}. This is not a surprise of course. Taking into account that the temperature is a measure of the mean energy per particle, and that time and energy are conjugated variables, it is clear to see how a redefinition of the former affects the latter. What thus fixes the right choice of foliation time? In order to answer this question, we must take into account that the rays that arrive to the asymptotic region of large radius have climbed the gravitational well by following free-falling trajectories, understood here as those ruled by their equations of motion\footnote{These rays do not follow geodesics of the metric, but they are nevertheless the natural trajectories with no external forces acting.}. Thus, we synchronize the clocks in the asymptotic region with that dictated by the rays. This is tantamount to select the proper vacuum state, and the right notion of energy for the field.

In order to make this connection explicit, we aim to compute the four-velocity of the ray close to the UH, following \cite{cro14}. We thus recall \eqref{kleinomega} and solve it together with the dispersion relation $\omega(k_\rho)$ to obtain the value of $k_\rho$. For this, it suffices to consider only the UV limit of the dispersion relation, since the modes that arrive at the asymptotic region correspond to hard blue-shifted modes close to the UH. Hence, we take $\omega= k_\rho^{Z}/\Lambda^{Z-1}$, and perform a WKB expansion for large momentum $k_\rho\sim \Lambda$. At leading order we get
\begin{align}\label{lambdaentw}
    k_\rho^{Z-1} \Lambda^{1-Z}= \frac{1-F(r)A^2(r)}{1+F(r)A^2(r)}.
\end{align}

From here we compute the group velocity of the field as
\begin{align}\label{ck}
    c(k_\rho)=\frac{{\rm d}\omega}{{\rm d} k_\rho}=Z\times \frac{1-F(r)A^2(r)}{1+F(r)A^2(r)},
\end{align}
and we use it to build the four-velocity
\begin{align}
    V^\m=U^\mu + c(k_\rho) S^\mu.
\end{align}

The vector $V^\m$ is then employed to extract the trajectory of the field. In particular we are interested in ${\rm d} v/{\rm d}r=V^0/V^1$, which close to the UH reads
\begin{align}\label{strahl}
\left.\frac{{\rm d} v}{{\rm d}r}\right|_{\scaleto{\rm UH}{3pt}}=-\frac{Z}{Z-1} \frac{B(r_{\scaleto{\rm UH}{3pt}})}{\sqrt{-F(r_{\scaleto{\rm UH}{3pt}})}}\frac{1}{N}+{\cal O}(N).
\end{align}

We can compare this with the relation ${\rm d}\tau/{\rm d}r$, which can be obtained by noting that close to the UH, $U_0\sim 0$ so that
\begin{align}
    -N {\rm d}\tau= \left. U_\m {\rm d}x^\mu\right|_{\scaleto{\rm UH}{3pt}}=U_r {\rm d}r ,
\end{align}
and therefore
\begin{align}
   \left. \frac{{\rm d}\tau}{{\rm d}r}\right|_{\scaleto{\rm UH}{3pt}}=-\frac{U_r}{N}=-\frac{B(r_{\scaleto{\rm UH}{3pt}})}{\sqrt{-F(r_{\scaleto{\rm UH}{3pt}})}}\frac 1 N.
\end{align}

Thus, we observe that the clock of the rays aligns with the foliation clock close to the UH, with a proportionality factor
\begin{align}
    \frac{{\rm d}\tau}{{\rm d}v}=\frac{Z-1}{Z},
\end{align}
which in turn implies that the ray sees a lapse
\begin{align}
    \hat N=\frac{Z-1}{Z} N,
\end{align}
close to the UH. Propagating this synchronization factor throughout our computation, we thus find
\begin{align}\label{eq:true_temp}
    \hat{T}_{\rm H}=\frac{Z-1}{Z}\frac{\kappa_{\scaleto{\rm UH}{3pt}}}{\pi},
\end{align}
which again agrees with previous results in the literature \cite{del22,he21}.

\section{Conclusions and discussion}\label{sec:conclusions}

In this work, we have studied the application of the gravitational tunneling method to derive the distribution of radiation emitted by a universal horizon in Einstein-Aether and Ho\v rava--Lifshitz Gravity. We have done it by focusing on spherically symmetric black hole solutions, endowed with a gravitating scalar field satisfying the symmetries of the background and with an anisotropic scaling between time and spatial directions. Our findings suggest that the collection of knowledge built from understanding the role of quantum fields close to the event horizon of a general relativistic black hole translates almost straightforwardly to the case under study here.

In particular, we have found that there exist non-vanishing tunneling probabilities associated to classically forbidden trajectories escaping the interior of the UH, which acts here as a universal causal boundary, a role which is again akin to the one of the event horizon in General Relativity. The classical action for these trajectories, which peel off the UH when traced back in time from the asymptotic region where our observer is placed, develops a non-vanishing imaginary part, inherited from a pole in the spatial momentum, which leads to a well-defined temperature once the tunneling probability is compared to a Boltzmann distribution, given by \eqref{eq:true_temp}. This can be interpreted as the existence of Hawking radiation emitted by the UH in a thermal ensemble made of quanta of the scalar field. The advantage of the tunneling method here with respect to other approaches is apparent. Through analyzing the behavior of the modes of the scalar fields close to the UH we have been able to obtain the value of the temperature for all possible static and spherically symmetric black hole solutions endowed with a UH. This is in agreement with recent works where this issue is analyzed from other directions.

An important property to highlight here is that the temperature shows a dependence on the dispersion relation of the scalar field through the pre-factor $(Z-1)/Z$, where $2Z$ is the highest power of the spatial preferred momentum in the square of the dispersion relation -- i.e. $\omega^2\sim k_\rho^{2Z}$ at large momentum. This factor has an important significance for model building and for the consistency of the model discussed here. Unless all matter species coupled to gravity share the same UV scaling in their dispersion relation, the temperatures of the different thermal ensembles emitted by the black hole will differ. In such a case, it is not difficult to envision how to construct a perpetuum mobile of the second kind, thus violating the laws of thermodynamics \cite{Dubovsky:2006vk}. Therefore, our results here suggest that a universal high energy behavior of the dispersion relation should exist in a consistent theory. This is in line with findings and discussions on the renormalization group flow properties of models coupling Ho\v rava--Lifshitz Gravity to matter actions, where indeed a universal UV scaling is present and controlled by the gravitational action \cite{Pospelov:2010mp}.

Some open questions linger though. In particular, and although we have determined the spectrum of radiation emitted by the UH, there is still a lack of profound understanding on the role of the Killing horizon in the process. Although no radiation is emitted by this surface in the present setting, as we have discussed in subsection \ref{sec:killing_radiation}, it might still have an important effect on the distribution of modes measured by an observer sitting at large radius, in the exterior of all horizons. Different rays of different energies will climb the gravitational well at a different rate, due to the momentum dependence of the dispersion relation and group velocity. In particular, rays with small momentum will linger for a long time close to the Killing horizon before being able to escape from it, accumulating energy and shifting their frequency. One could then expect a distortion of the spectrum measured by an asymptotic observer when compared to the emitted one
\begin{align}
    \rho_{\rm measured}(\Omega)=\int \frac{{\rm d}\Omega'}{2\pi}\  {\cal G}(\Omega,\Omega')\times \rho_{\rm emitted}(\Omega'),
\end{align}
where the graybody kernel ${\cal G}(\Omega,\Omega')$ encodes the effects introduced by the Killing horizon. Its computation is beyond the scope of this work, but is poses an exciting problem for future research. Is the measured spectrum still the one of a thermal ensemble with temperature $\hat T_{\rm UH}$? Is it shifted so that it looks like a thermal emission with $T_{\rm H}$ instead, thus washing off the presence of Lorentz violations? Or is it something in between?

Finally, it is worth to mention that all of our results here are obtained by assuming a fixed background with no back-reaction onto the geometry due to the presence of the matter fields. In particular, the black holes discussed here are solutions only to the \emph{low energy} Lagrangian of HL Gravity -- usually denoted by ${\cal L}_2$ -- and described here in terms of EA gravity, which serves as an effective field theory description of Lorentz violations at the second derivative level. 
However, the more realistic -- and perhaps UV complete \cite{Barvinsky:2017kob,Barvinsky:2019rwn,Barvinsky:2021ubv} -- action of HL Gravity contains also terms with four and six derivatives -- in $3+1$ dimensions. Taking into account the fact that the modes investigated here, and ultimately responsible of the Hawking radiation, are blue-shifted to high energies, one could wonder if, in a general case, they would trigger perturbations of the geometry with comparable momentum, thus exiting the applicability of the low energy theory ${\cal L}_2$. This would demand the presence of the higher order terms in order to account for their dynamics. Stability of the UH, or even of the full space-time, in the presence of these perturbations is so far an open problem in the field; and one that we should keep in mind when discussing scenarios like the one in this work.

\acknowledgements
M. H-V. wants to thank the APP department at SISSA for their hospitality during the final stages of this work. The work of F. D. P., S. L., and M. S. has been supported by the Italian Ministry of
Education and Scientific Research (MIUR) under the grant PRIN MIUR 2017-MB8AEZ. The work of M. H-V. has been supported by the Spanish State Research Agency MCIN/AEI/10.13039/501100011033 and by the EU NextGenerationEU/PRTR funds, under grant IJC2020-045126-I. IFAE is partially funded by the CERCA program of the Generalitat de Catalunya.

\bibliography{littunnel}{}
\end{document}